\documentclass[conference]{IEEEtran}
\IEEEoverridecommandlockouts
\usepackage{cite}
\usepackage[utf8]{inputenc}
\usepackage{array}
\usepackage{adjustbox}
\usepackage{makecell}
\usepackage[framemethod=TikZ]{mdframed}
\usepackage{amsmath,amssymb,amsfonts}
\usepackage{graphicx}
\usepackage{lipsum} % For generating dummy text
\usepackage{textcomp}
\usepackage{float}
\usepackage{algorithm}
\usepackage{multirow}
\usepackage{algpseudocode}

\usepackage{cuted} % for the strip environment
\usepackage[margin=0.6in]{geometry}
\usepackage{tcolorbox}
\tcbuselibrary{skins,breakable}

\newtcolorbox{QA}[2][]{
  breakable,
  colframe = black,
  colback  = white,
  coltitle = black,
  fonttitle=\bfseries,
  title    = #2,
  #1,
}

\newtcolorbox{promptanswerbox}{
    colback=gray!5, % Background color of the box
    colframe=black!75, % Frame color
    coltitle=black, % Title text color
    title=Prompt, % Title of the box
    fonttitle=\bfseries, % Bold title font
    attach title to upper, % Attach the title to the content
    segmentation style={black!75, dashed}, % Style of the separator line
}
\usepackage{array} % For table formatting
\usepackage{booktabs} % For formal tables
\usepackage{algpseudocode}
\usepackage{xcolor}
\def\BibTeX{{\rm B\kern-.05em{\sc i\kern-.025em b}\kern-.08em
    T\kern-.1667em\lower.7ex\hbox{E}\kern-.125emX}}
\usepackage{comment}
\usepackage{makecell}
\usepackage{amsmath}
\usepackage{amsmath, amssymb, amscd, amsthm, amsfonts}
\usepackage{colortbl}
\usepackage{soul}
\usepackage{multicol}

\usepackage[colorlinks=true,linkcolor=blue,anchorcolor=black,citecolor=red,filecolor=black,menucolor=black,runcolor=black,urlcolor=black]{hyperref}
    
\begin{document}

\title{%Exploiting Task-Oriented Dialogue in GenAI Models for Anomaly Detection in Digital Substations
Leveraging Conversational Generative AI for Anomaly Detection in Digital Substations}

\author{Aydin Zaboli, \IEEEmembership{Student Member, IEEE}, Seong Lok Choi, \IEEEmembership{Member, IEEE}, Junho Hong, \IEEEmembership{Senior Member, IEEE}
%\thanks{This paragraph of the first footnote will contain the date on which you submitted your paper for review. It will also contain support information, including sponsor and financial support acknowledgment. For example, ``This work was supported in part by the U.S. Department of Commerce under Grant BS123456.'' }
%\thanks{The next few paragraphs should contain the authors' current affiliations, including current address and e-mail. For example, F. A. Author is with the National Institute of Standards and Technology, Boulder, CO 80305 USA (e-mail: author@boulder.nist.gov). }
\thanks{A. Zaboli and J. Hong are with the Department of Electrical and Computer Engineering, University of Michigan -- Dearborn, Dearborn, MI 48128, USA.}
\thanks{S. L. Choi is with the Power Systems Engineering Center, National Renewable Energy Laboratory (NREL), Golden, CO 80401, USA.}
%\thanks{T. C. Author is with the Electrical Engineering Department, University of Colorado, Boulder, CO 80309 USA, on leave from the National Research Institute for Metals, Tsukuba, Japan (e-mail: author@nrim.go.jp).}
}

\maketitle

\begin{abstract}
This study addresses critical challenges of cybersecurity in digital substations by proposing an innovative task-oriented dialogue (ToD) system for anomaly detection (AD) in multicast messages, specifically, generic object oriented substation event (GOOSE) and sampled value (SV) datasets. Leveraging generative artificial intelligence (GenAI) technology, the proposed framework demonstrates superior error reduction, scalability, and adaptability compared with traditional human-in-the-loop (HITL) processes. Notably, this methodology offers significant advantages over machine learning (ML) techniques in terms of efficiency and implementation speed when confronting novel and/or unknown cyber threats, while also maintaining model complexity and precision. The research employs advanced performance metrics to conduct a comparative assessment between the proposed AD and HITL-based AD frameworks, utilizing a hardware-in-the-loop (HIL) testbed for generating and extracting features of IEC61850 communication messages. This approach presents a promising solution for enhancing the reliability of power system operations in the face of evolving cybersecurity challenges.
\end{abstract}


\section{Introduction} \label{intro}
%%%%%%%%%%%%%%%%%%%%%%%%%%%%%%%%%%%%%%%%%%%%%%
The integration of IEC 61850-based digital substations into smart grids has revolutionized energy management, enabling automated data collection and remote control of electrical systems. However, the fusion of power infrastructure with communication networks has introduced various security vulnerabilities, requiring the implementation of robust anomaly detection systems (ADSs)
%. The adoption of IEC 61850 and associated communication protocols, such as GOOSE and SV, has given rise to specialized malicious strategies with distinct traffic and attack configurations. Consequently, there is an urgent need for ADSs to acquire new signatures for training, testing, and evaluation purposes in order 
to effectively address challenges concerning the preservation and defense of critical national assets~\cite{
%khaw2020deep, 
zhu2020intrusion, hong2017intelligent, hong2022automated}.

While ML techniques have become instrumental in detecting anomalies within GOOSE and SV multicast messages, they face limitations in terms of scalability, decision-making efficacy, and data processing. The continuous need for model re-training to address new attack vectors creates temporal vulnerabilities and resource-intensive processes. Given these complexities and limitations, GenAI tools present a more flexible and adaptive approach to AD in digital substations~\cite{mohammadabadi2024generative}. GenAI, with its inherent capability to understand situational intricacies and subtleties, can potentially detect novel attacks without prior training, offering a more robust and efficient AD methodology. This innovative approach emphasizes the potential of GenAI tools to enhance cybersecurity by providing a tool that can dynamically evolve in response to emerging cyber threats, thereby addressing the limitations of traditional ML frameworks and HITL processes~\cite{mavikumbure2024generative, zaboli2024chatgpt1}.
%%%%%%%%%%%%%%%%%%% Literature Survey%%%%%%%%%%%%%%%%
Recent advancements in AD processes for securing digital substations have seen a significant shift toward the utilization of sophisticated ML models. These models have demonstrated considerable potential in enhancing AD capabilities by analyzing patterns and anomalies within datasets, thereby enabling real-time detection of cyberattacks and contributing to grid security~\cite{wang2022anomaly
%,reda2021vulnerability1
}. Notable among these efforts is the AI-based ransomware detection approach proposed by Alvee \textit{et al.}~\cite{alvee2021ransomware}, that employs a convolutional neural network (CNN) and innovatively converts binary files into 2-D image files for detection, achieving a high accuracy rate of $96.22\%$. However, the approach's efficacy remains subject to debate due to the absence of HIL testbed data and the absence of various attack scenarios.
%in the experimental design.
Further contributions to this field include the research of
%Upadhyay \textit{et al.}'s gradient boosting method for feature selection in IDSs, which shows promise in improving accuracy and reducing false positives (FPs) in the AD process. Nevertheless, their use of a general attack vector for performance analysis limits the differentiation between various types of attacks~\cite{upadhyay2020gradient}. 
%Nhung-Nguyen \textit{et al.}~\cite{nhung2024machine}, which proposed a deep neural network (DNN)-based ADS for detecting malicious GOOSE communications, demonstrating high accuracy but requiring large amounts of labeled training data. While these approaches have shown effectiveness, they face challenges in adapting to new attack types without retraining. 
Yang \textit{et al.}~\cite{yang2022new} who introduced a novel methodology combining statistical analysis and ML models for AD processes, offering improved adaptability to evolving threats.
Additionally, Zhu \textit{et al.}~\cite{zhu2020intrusion} have developed a novel ADS specifically targeting manufacturing message specification (MMS)-based measurement attacks, incorporating advanced detection algorithms to enhance accuracy. These diverse approaches highlight the ongoing efforts to refine and optimize AD techniques in the context of digital substation security, although each methodology presents its own set of limitations and areas for further research and development.
%%%%%%%%%%%%%%%%%%%%%%%%%% Justification %%%%%%%%%%%%%%%
Current literature reviews indicate a notable absence of research exploring directed methodologies leveraging GenAI tools for attack and anomaly detection based on human recommendations. Existing research endeavors predominantly grapple with challenges in scalability, adaptability, robustness, and processing efficiency. Consequently, there is a pressing need for a framework that can address challenges with minimal effort and reduced reliance on continuous human expert intervention.
%%%%%%%%%%%%%%%%%%%%%%%%%%%%%%%%%%%%%%%%%%%%%%%%%%%%%%%%
%\subsection{Contributions} \label{contributions}

This study presents an advanced GenAI-based ToD system, 
%(e.g., Anthropic Claude Pro~\cite{anthropic}), 
offering advantages over HITL processes and ML algorithms for AD in multicast messages. Built on historical human recommendations, it automates decision-making by emulating patterns, potentially reducing errors over time. The learning capabilities and data processing proficiency enable scalability, though challenges remain in building user trust.
%and handling very complex situations. 
The following points encapsulate contributions of the proposed approach:
\begin{itemize}
    \item This paper introduces a groundbreaking implementation of a GenAI-based ToD system for the efficient and reliable detection of anomalies in multicast messages. This innovative approach presents a method for bolstering the security and operational stability of smart grid infrastructures.
    \item An analysis of the proposed methodology was conducted employing a diverse array of advanced metrics. This rigorous evaluation process not only rectifies the limitations identified in prior research but also institutes a new paradigm for assessing ADS efficacy with improved efficiency, flexibility, and expandability.
\end{itemize}

%\subsection{Paper Layout}
The subsequent sections are organized thusly: An assessment of IEC 61850-based protocols, and human-derived rules is presented in Section~\ref{cyberintro}. Section~\ref{proposedGenAIToD} elucidates the proposed AD methodology in GOOSE/SV datasets. Section~\ref{results&discussion} encompasses a detailed discussion of results, conducting a comparative analysis of the efficacy of the HITL approach with that of the proposed framework. Finally, this research concludes in Section~\ref{conclusion}. 
%with a synthesis of key findings and suggestions for future investigative directions.

\section{Substation Cyber Imperatives} \label{cyberintro}
The integration of advanced communication technologies into digital substations necessitates robust cybersecurity measures, encompassing multi-layered protective strategies and regular assessments to address evolving cyber threats. An HIL testbed provides a controlled environment for investigating the interplay between cyber breaches and power system resilience. 
%This testbed incorporates diverse components, including intelligent electronic devices (IEDs), software-defined networking (SDN) switches, and supervisory control and data acquisition (SCADA) systems, synchronized via GPS. The architecture facilitates real-time analysis of communication dynamics and information processing, crucial for enhancing ADS and mitigation strategies. 
%The proposed GenAI-based framework is designed to detect anomalies within multicast messages, leveraging connectivity with SDN switches for comprehensive protection~\cite{reda2021vulnerability1, hong2017intelligent}. 
GOOSE and SV packet extraction from the HIL testbed is executed using Wireshark, enabling comprehensive network traffic analysis. This process facilitates detailed observation of communication patterns within the testbed environment. This methodical approach ensures accurate data collection and provides critical insights into cyber-physical system (CPS) dynamics~\cite{zaboli2024chatgpt1}. The next section describes datasets, their feature extraction process and human recommendations integral to the proposed framework.
%%%%%%%%%%%%%%%%%%%%%%%%%%%%%%%%%%%%%%%%%%%%%%%%%%%%%%%%
\subsection{GOOSE and SV Dataset Features \& Rules}
The most important features of GOOSE messages can be considered as time, destination MAC address (\textit{DM}) ($01$ $00$ $03$), source MAC address (\textit{SM}) ($27$ $34$ $31$), \textit{type} ($88b8$), application identifier (\textit{appid}) ($3$), \textit{dataset} ($SEL\_421\_1CFG/LLN0\$Goose$), GOOSE identifier (\textit{goid}) ($SEL\_421\_1$), state number (\textit{stnum}) ($27$), sequence number (\textit{sqnum}), and \textit{data1}/\textit{data2} values that are binary. Let G = (\textit{time}, \textit{DM}, \textit{SM}, \textit{type}, \textit{appid}, \textit{dataset}, \textit{goid}, \textit{stnum}, \textit{sqnum}, \textit{data1}/\textit{data2}) represent the features of a GOOSE message. Eqs.~(\ref{GR1})--(\ref{GR8}) illustrate the GOOSE rules (i.e., \textit{GR\#1} to \textit{GR\#8}) employed in this paper to check the different abnormalities of datasets. This paper presents different levels of considering rules as without training, WT (i.e, no rules), partial training, PT (i.e., \textit{GR\#1} to \textit{GR\#5}), and full training, FT (i.e., \textit{GR\#1} to \textit{GR\#8}). 
%from \textit{GR\#1} to \textit{GR\#8}, respectively.
\textit{GR\#1}: When consecutive data packets exhibit identical \textit{DM} and \textit{SM} attributes, the \textit{sqnum} parameter should incrementally advance. Any inconsistency reveals an abnormality.
\begin{equation} \label{GR1}
\small
GR\#1(G_i, G_{i-1}) = 
\begin{cases}
    1, & \text{if } DM_i = DM_{i-1} \land SM_i = SM_{i-1} \land \\ 
       & \quad sqnum_i = sqnum_{i-1} + 1 \\
    0, & \text{otherwise}
\end{cases}
\end{equation}
\textit{GR\#2}: Modifications in \textit{data1} (\textit{d1}) or \textit{data2} (\textit{d2}) necessitate a unitary increment in \textit{stnum} and an \textit{sqnum} reset to $0$. Non-compliance with this protocol indicates an anomaly.
\begin{equation} \label{GR2}
\small
GR\#2(G_i, G_{i-1}) = 
\begin{cases}
    1, & \text{if } (d1_i \neq d1_{i-1} \lor d2_i \neq d2_{i-1}) \land stnum_i \\ 
       & = stnum_{i-1} + 1 \land sqnum_i = 0 \\
    0, & \text{otherwise}
\end{cases}
\end{equation}
\textit{GR\#3}: For data with identical \textit{DM} and \textit{SM}, the \textit{stnum} must maintain a monotonically increasing sequence. Any regression in \textit{stnum} value constitutes an anomaly.
%\begin{figure*}[!h]
\begin{equation} \label{GR3}
\small
    GR\#3(G_i, G_{1:i-1}) = 
    \begin{cases}
        1, & \text{if } DM_i = SM_i \land stnum_i > \\ & \max_{j<i}(stnum_j) \\
        0, & \text{otherwise}
    \end{cases}
\end{equation}
%\end{figure*}
\textit{GR\#4}: Any alteration in \textit{DM}, \textit{SM}, \textit{type}, \textit{appid}, \textit{dataset}, or \textit{goid} parameters is indicative of an anomalous condition.
%----------------------------------------------
\begin{equation} \label{GR4}
\small
GR\#4(G_i, G_{i-1}) = 
\begin{cases}
    1, & \text{if } DM_i = DM_{i-1} \land SM_i = SM_{i-1} \land \\ 
       & \quad type_i = type_{i-1} \land dataset_i = dataset_{i-1} \\
    0, & \text{otherwise}
\end{cases}
\end{equation}
%----------------------------------------------
\textit{GR\#5}: The \textit{time} column must adhere to a format delineating hour, minute, and second with microsecond precision. Any deviation from this temporal feature constitutes an anomaly.
%\begin{figure}[!h]
\begin{equation} \label{GR5}
\small
    GR\#5(time_i) = 
    \begin{cases}
        1, & \text{if } time_i \text{ is in format HH:MM:SS.mmmmmm} \\
        0, & \text{otherwise}
    \end{cases}
\end{equation}
%\end{figure}
\textit{GR\#6}: The occurrence of data frequency exceeding $10$ instances within a $10$ $\mu$s interval is classified as an anomalous event.
%\begin{figure}[!h]
\begin{equation} \label{GR6}
\small
    GR\#6(G_{i-9:i}) = 
    \begin{cases}
        1, & \text{if } \forall j \in [i-9, i-1]: time_{j+1} - time_j \leq 10\mu s \\
        0, & \text{otherwise}
    \end{cases}
\end{equation}
%\end{figure}
\textit{GR\#7}: A temporal gap in data transmission exceeding $10$ seconds is indicative of an anomalous condition.
%\begin{figure}[!h]
\begin{equation} \label{GR7}
\small
    GR\#7(G_i, G_{i-1}) = 
    \begin{cases}
        1, & \text{if } time_i - time_{i-1} \leq 10s \\
        0, & \text{otherwise}
    \end{cases}
\end{equation}
%\end{figure}
\textit{GR\#8}: Upon detection of alterations in \textit{d1} or \textit{d2}, the \textit{stnum} should remain constant while the \textit{sqnum} undergoes an increment. Any deviation depicts an anomaly.
%\begin{figure*}[!h]
\begin{equation} \label{GR8}
\small
    GR\#8(G_i, G_{i-1}) = 
    \begin{cases}
        1, & \text{if } (d1_i \neq d1_{i-1} \lor d2_i \neq d2_{i-1}) \land stnum_i = \\ & stnum_{i-1} \land sqnum_i > sqnum_{i-1} \\
        0, & \text{otherwise}
    \end{cases}
\end{equation}
%\end{figure*}
In the case of SV datasets, the most important features can be denoted as \textit{time}, \textit{DM} ($04$ $00$ $01$), \textit{SM} ($27$ $22$ $13$), \textit{type} ($88ba$), \textit{appid} ($40$), sampled value identifier (\textit{svid}) ($4000$), and sample count (\textit{smpcnt}). Let S = (\textit{time}, \textit{DM}, \textit{SM}, \textit{type}, \textit{appid}, \textit{svid}, \textit{smpcnt}) represent the features of an SV message. Eqs.~(\ref{SR1})--(\ref{SR8}) illustrate the SV rules (i.e., \textit{SR\#1} to \textit{SR\#8}) utilized to check the anomalies of SV datasets. A similar procedure of rules for different level of training in GOOSE messages can be applied to SV messages.
%from \textit{SR\#1} to \textit{SR\#8}, correspondingly.
\textit{SR\#1}: The \textit{smpcnt} parameter is constrained to the integer interval [$0, 4799$] for $60$ Hz systems (i.e., $80 \times 60=4800$) or [$0, 3999$] for $50$ Hz systems (i.e., $80 \times 50=4000$). Any value outside this prescribed range is classified as an anomalous condition.
%\begin{figure}[!h]
\begin{equation} \label{SR1}
\small
    SR\#1(S_i) = 
    \begin{cases}
        1, & \text{if } 0 \leq Smpcnt_i \leq 4799 \\
        0, & \text{otherwise}
    \end{cases}
\end{equation}
%\end{figure}
\textit{SR\#2}: The \textit{smpcnt} parameter should exhibit a uniformly increasing series from $0$ to $4799$ for $60$ Hz systems, followed by a reset to $0$. Any deviation from this progression is indicative of an anomaly.
\begin{equation} \label{SR2}
\resizebox{1.0\hsize}{!}{$
\scriptsize
SR\#2(S_i, S_{i-1}) = 
\begin{cases}
    1, & \text{if } (Smpcnt_i > Smpcnt_{i-1} \land \\ & Smpcnt_i \leq 4799) \lor \\ 
       & \quad (Smpcnt_i = 0 \land Smpcnt_{i-1} = 4799) \\
    0, & \text{otherwise}
\end{cases}
$}
\end{equation}
\textit{SR\#3}: The \textit{smpcnt} parameter must maintain a non-decreasing sequence until it attains the value $4799$, whereupon it resets to $0$. Any deviation from this pattern denotes an anomaly.
%\begin{figure*}[!h]
\begin{equation} \label{SR3}
\resizebox{1.0\hsize}{!}{$
\small
    SR\#3(S_i, S_{i-1}) = 
    \begin{cases}
        1, & \text{if } Smpcnt_i \geq Smpcnt_{i-1} \lor \\ & (Smpcnt_i = 0 \land Smpcnt_{i-1} = 4799) \\
        0, & \text{otherwise}
    \end{cases}
$}
\end{equation}
%\end{figure*}
\textit{SR\#4}: The parameters \textit{DM}, \textit{SM}, \textit{type}, \textit{appid}, and \textit{svid} must maintain invariance across all conditions. Any changes in these values signifies an anomalous state.
%\begin{figure*}[!h]
\begin{equation} \label{SR4}
\resizebox{1.0\hsize}{!}{$
\small
    SR\#4(S_i, S_{i-1}) = 
    \begin{cases}
        1, & \text{if } DM_i = DM_{i-1} \land SM_i = SM_{i-1} \land \\ & type_i = type_{i-1} \land apid_i = apid_{i-1} \land \\ & svid_i = svid_{i-1} \\
        0, & \text{otherwise}
    \end{cases}
$}
\end{equation}
%\end{figure*}
\textit{SR\#5}: The temporal data field must conform to a hierarchical structure of hour, minute, second, and microsecond. A deviation from this chronological format indicates an anomaly.
%\begin{figure}[!h]
\begin{equation} \label{SR5}
\small
    SR\#5(time_i) = 
    \begin{cases}
        1, & \text{if } time_i \text{in format HH:MM:SS.mmm} \\
        0, & \text{otherwise}
    \end{cases}
\end{equation}
%\end{figure}
\textit{SR\#6}: The standard temporal interval is constrained to the range of $200$ to $215$ $\mu$s. Any deviation from this prescribed timeframe indicates an anomalous condition.
%\begin{figure}[!h]
\begin{equation} \label{SR6}
\small
    SR\#6(S_i, S_{i-1}) = 
    \begin{cases}
        1, & \text{if } 200\mu s \leq time_i - time_{i-1} \leq 215\mu s \\
        0, & \text{otherwise}
    \end{cases}
\end{equation}
%\end{figure}
\textit{SR\#7}: The occurrence of data frequency exceeding $12$ instances within a $2.083$-$ms$ interval is an anomalous event.
%\begin{figure}[!h]
\begin{equation} \label{SR7}
\small
    SR\#7(S_{i-11:i}) = 
    \begin{cases}
        1, & \text{if } time_i - time_{i-11} \leq 2.083ms \\
        0, & \text{otherwise}
    \end{cases}
\end{equation}
%\end{figure}
\textit{SR\#8}: The \textit{smpcnt} parameter should exhibit a unitary increment with each successive instance. Any irregularity in this sequential progression implies an anomalous condition.
%\begin{figure}[!h]
\begin{equation} \label{SR8}
\small
    SR\#8(S_i, S_{i-1}) = 
    \begin{cases}
        1, & \text{if } Smpcnt_i = (Smpcnt_{i-1} + 1) \\
        0, & \text{otherwise}
    \end{cases}
\end{equation}
%\end{figure}
The proposed framework is crafted in which GOOSE (\textit{GR\#1} to \textit{GR\#8}) and SV (\textit{SR\#1} to \textit{SR\#8}) rules guide SQL queries that filter messages to detect anomalies. In general, this framework uses partial training to help analysts apply rules for identifying abnormal patterns, suggesting additional rules if no anomalies are found. In full training, this method tries to confirm anomalies, by analyzing recent message packets and refining detection through belief-action updates. This adaptive approach enhances AD accuracy by dynamically incorporating rule adjustments which will be elaborated in the next section.
%%%%%%%%%%%%%%%%%%%%%%%%%%%%%%%%%%%%%%%%%%%%%%%%%%%%%%%%
\section{A Novel Anomaly Detection System Paradigm: Integrating GenAI with Task-Oriented Dialogue} \label{proposedGenAIToD}
\begin{figure}[!h]
\centerline{\includegraphics[width=0.8\columnwidth]{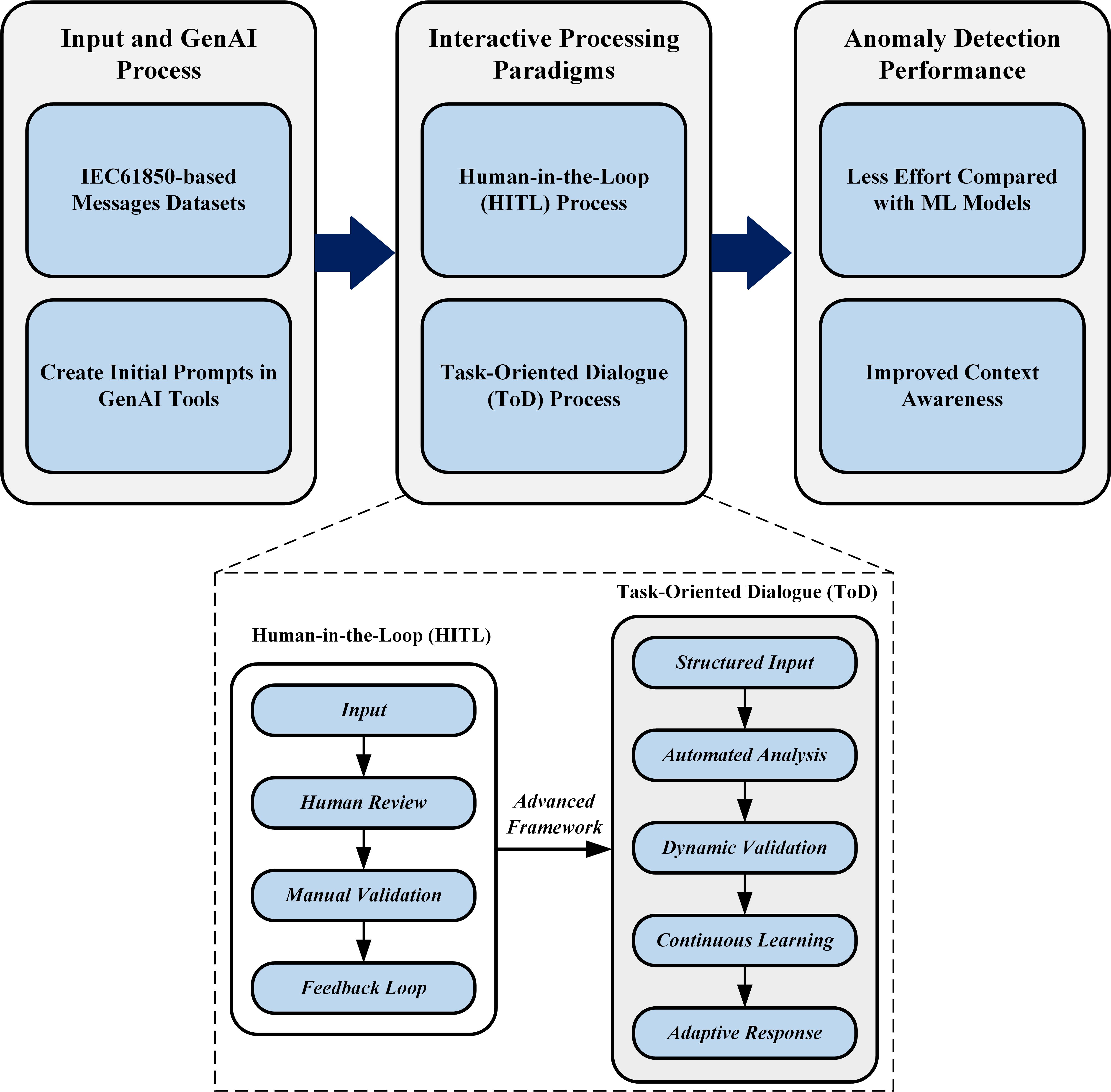}}
\caption{A general framework of interactive GenAI-based AD process.}
\label{fig:GenAI}
\end{figure}

\begin{figure*}[!t]
\centerline{\includegraphics[width=1.6\columnwidth]{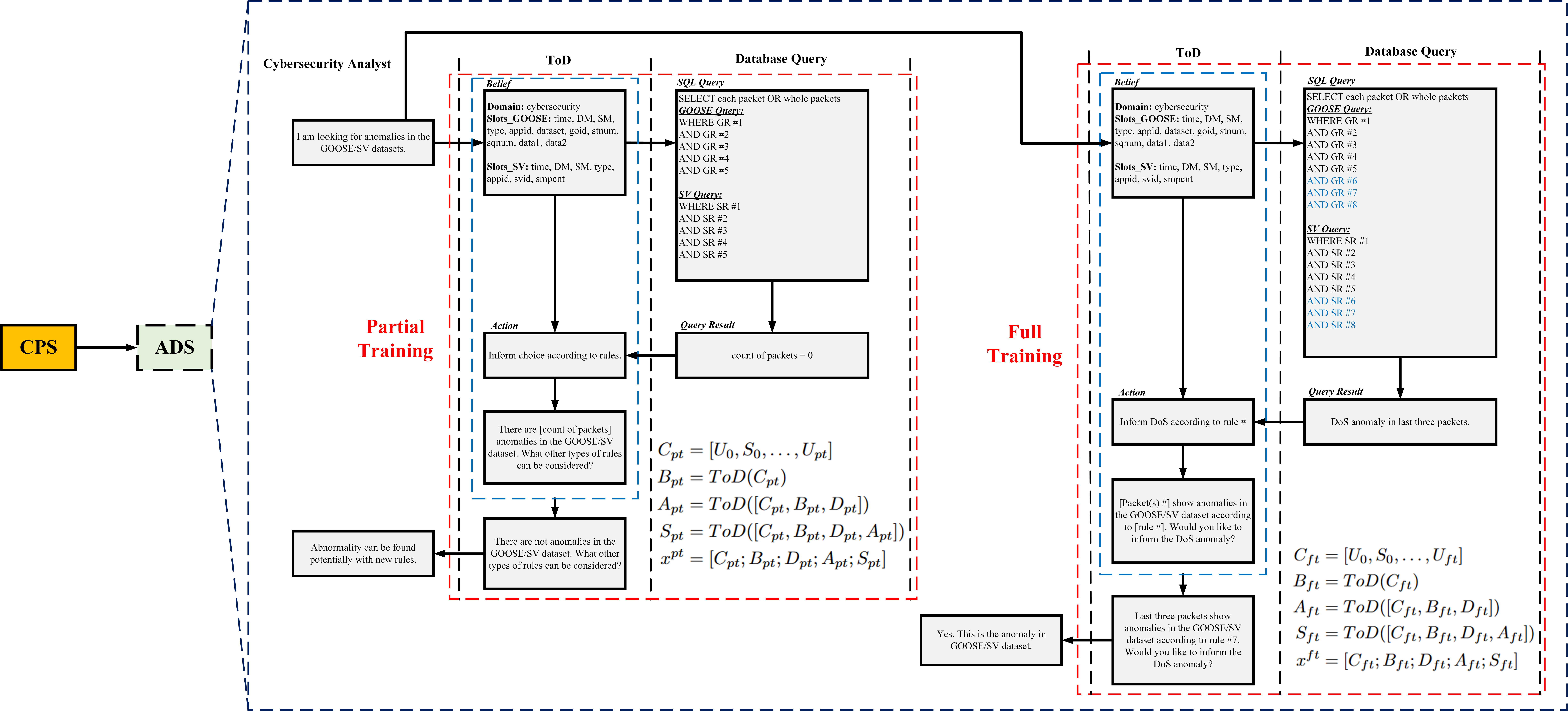}}
\caption{Advancing multicast messages security: A GenAI-based ToD System for the AD process.}
\label{fig:ToDframework2025}
\end{figure*}

%-----------------------------------------
A GenAI-driven approach to AD in GOOSE and SV datasets can enhance the detection process, using interactive processing paradigms such as HITL and ToD as demonstrated in Fig.~\ref{fig:GenAI}.
The framework begins with gathering IEC61850-based message datasets. Initial prompts are crafted in GenAI tools to help the system identify anomalies by establishing normal operational patterns and highlighting potential threats. The interactive processing then enhances this process through HITL and ToD which the HITL process integrates human feedback to iteratively refine the model's detection accuracy, while ToD allows the GenAI system to engage in dialogue, gathering context and adjusting its response. These steps create a flexible, adaptable detection system that can efficiently identify anomalies. Compared to the ToD approach, the HITL process exhibits several limitations. While HITL relies heavily on constant human review and manual validation, requiring time and effort for each feedback loop, ToD offers a more efficient automated analysis system with dynamic validation capabilities. The HITL process's dependence on human intervention creates scalability constraints and slower processing cycles. Different steps of the ToD system are organized below to show its outperformance on HITL processes and ML models.
%-----------------------------------------
ToD systems' efficacy is traditionally evaluated through their proficiency in distinct subtasks, encompassing dialogue state tracking (also known as belief state management), dialogue management (which includes action and decision prediction), and the generation of responses utilizing an SQL query database. In this GenAI-based framework, SQL query block acts as a filtering tool that efficiently identifies anomalies in data by checking GOOSE and SV messages against predefined rules. This approach simplifies the AD by processing only relevant data and maintaining consistent rule application to detect issues. This partitioning of tasks has facilitated the development of specialized models for each subtask, a methodology that has gained considerable adoption within the field~\cite{hu2024dialight}. The present research endeavors to investigate the efficacy of a unified and end-to-end model in managing these multilayered functions, as illustrated in Fig.~\ref{fig:ToDframework2025} along with CPS. This model incorporates a cybersecurity analyst component, implemented as a GenAI tool, which processes GOOSE and SV data to detect anomalies, leveraging packets, ToD labels, and anomaly scores as inputs to foster an understanding of anomalous characteristics.
%%%%%%%%%%%%%%%%%%%%%%%%%%%%%%%%%%%%%%%%%%%%
%----------------------------------------------
%Within the framework of dialogues, each interaction is conceptualized as a series of consecutive turns, where at a given turn ``$t$,'' the user input (denoted as $U_t$) infers a system response ($S_t$). The ToD model, particularly during the inference phase, assimilates all preceding dialogue turns as contextual information ($C_t = [U_0; S_0; ...; U_t]$), subsequently generating a belief state ($B_t$) at each turn. 
%This belief state comprises triplets that encapsulate specific values for various slots within the ``cybersecurity'' domain, structured as (domain, slot name, value), with distinct slots for GOOSE and SV multicast messages and their respective features. The model then determines an action ($A_t$) based on the contextual information, belief state, and domain, producing a result ($x^t$) that encompasses various parameters integral to the proposed framework. This framework, in conjunction with predefined human recommendations, is embodied within the GenAI tool to facilitate the AD process in datasets. The SQL query block, which contains diverse human-derived recommendations, is classified into partial and full training levels, with 5 and 8 recommendations respectively, thereby enabling the presentation of results through an iterative loop of directed dialogues.
%------------------------------------------------
A separation of the proposed ToD framework's steps illustrated in Fig.~\ref{fig:GenAI} and elaborated below.

\textbf{Structured Input:} This block captures structured inputs essential for the ToD system to function effectively which corresponds to data parameters (e.g., \textit{time}, \textit{DM}). For PT and FT levels, input packets are represented as $C_{pt} = [U_0, S_0, \dots, U_{pt}]$ and $C_{ft} = [U_0, S_0, \dots, U_{ft}]$, respectively, which denotes a collection of structured data extracted from GOOSE/SV packets.

\textbf{Automated Analysis:} This step involves analyzing data automatically, applying AD rules based on prior rules. It includes SQL queries for messages to extract anomaly-related data. During the PT level, automated analysis is stated as $B_{pt} = \text{ToD}(C_{pt})$ which indicates the transformation of the structured input to identify preliminary anomalies based on rules. Then, the analysis is refined in the FT level as $B_{ft} = \text{ToD}(C_{ft})$, providing a deeper analysis for complete training.

\textbf{Dynamic Validation:} This step includes verifying identified anomalies dynamically, using a set of conditions. It includes queries to count specific packet conditions. In the PT phase, dynamic validation refines the analysis with additional packet transformations as $S_{pt} = \text{ToD}([C_{pt}, B_{pt}, D_{pt}, A_{pt}])$ where \( D_{pt} \) includes additional conditions. Further, the validation includes more advanced criteria in the FT phase as $S_{ft} = \text{ToD}([C_{ft}, B_{ft}, D_{ft}, A_{ft}])$, increasing the complexity and robustness of validation checks with further transformations.

\textbf{Continuous Learning:} The framework adapts over time, incorporating feedback to refine the AD process. This aligns with the transition from partial to full training levels in the proposed ToD framework. In this case, the learning happens iteratively in PT with \( A_{pt} \), where rules are revised to enhance AD as $x^{pt} = [C_{pt}; B_{pt}; D_{pt}; A_{pt}; S_{pt}]$. Moreover, the learning process is formalized with a higher count of structured feedback and adaptive rules in FT level as $x^{ft} = [C_{ft}; B_{ft}; D_{ft}; A_{ft}; S_{ft}]$.

\textbf{Adaptive Response:} Based on AD outcomes, the system generates adaptive responses to inform the cybersecurity analyst of potential issues in the GOOSE/SV data. The responses align with the system's blocks for communicating anomaly findings. The adaptive responses for PT and FT levels involve generating responses based on the findings in $x^{pt}$ and $x^{ft}$, respectively. 
%Each of these steps leverages the custom ToD framework's specific query results, structured inputs, and domain knowledge of cybersecurity to detect and respond to anomalies effectively in digital substations. This approach ensures continuous improvement from partial to full training and accurate anomaly detection to safeguard the system.

%-----------------------------------------------
%\subsection{ML and HITL Paradigms: Key Obstacles}
An evaluation of contemporary AD paradigms, including HITL processes, ML architectures, and the novel GenAI-based ToD framework, illuminates the efficacy and versatility of the latter approach. While HITL methodologies are inherently limited by the availability and expertise of human operators, 
%(potentially leading to decision-making inconsistencies)
and ML models, despite their computational capability, are constrained by input/output structures and reduced adaptability, 
%(necessitating recurrent training for novel scenarios)
this system emerges as a more comprehensive and efficient solution. This innovative framework harnesses the power of natural language dialogue to accommodate complex queries and responses, demonstrating remarkable adaptability to emergent scenarios and anomalies through advanced prompt engineering techniques, while concurrently offering interpretable insights. 
%The proposed system's capacity for intuitive linguistic interactions diminishes the necessity for extensive operator training, its proficiency in processing and analyzing unstructured textual data renders it particularly adept at handling intricate datasets, and its inherent scalability coupled with efficient adaptation to novel threats through continuous learning and fine-tuning mechanisms ensure robust AD capabilities~\cite{chung2023instructtods}. 
%These diverse advantages, including the framework's orientation for rapid adaptation and its ability to effectively navigate complex scenarios, collectively establish this ToD approach as an enhanced alternative to traditional HITL and ML methods in the domain of cybersecurity.
%%%%%%%%%%%%%%%%%%%%%%%%%%%%%%%%%%%%%%%%%%%%%%%%%%
\section{Results and Discussion} \label{results&discussion}
\begin{figure}[!h]
\centerline{\includegraphics[width=0.9\columnwidth]{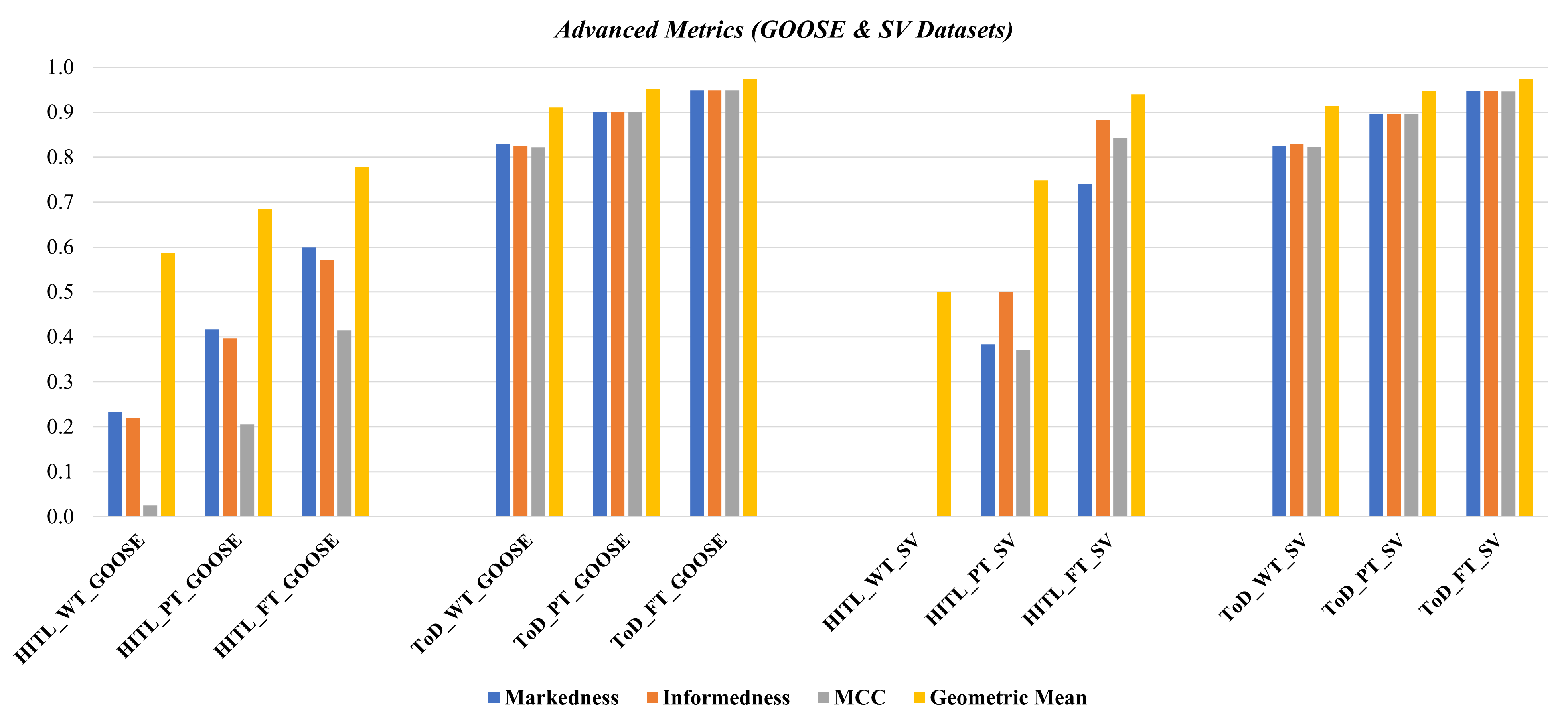}}
\caption{A comparative assessment of GenAI-based HITL and the proposed framework: Insights from advanced metrics applied to GOOSE and SV datasets.}
\label{MetricAll}
\end{figure}

\begin{figure}[!h]
\centerline{\includegraphics[width=0.8\columnwidth]{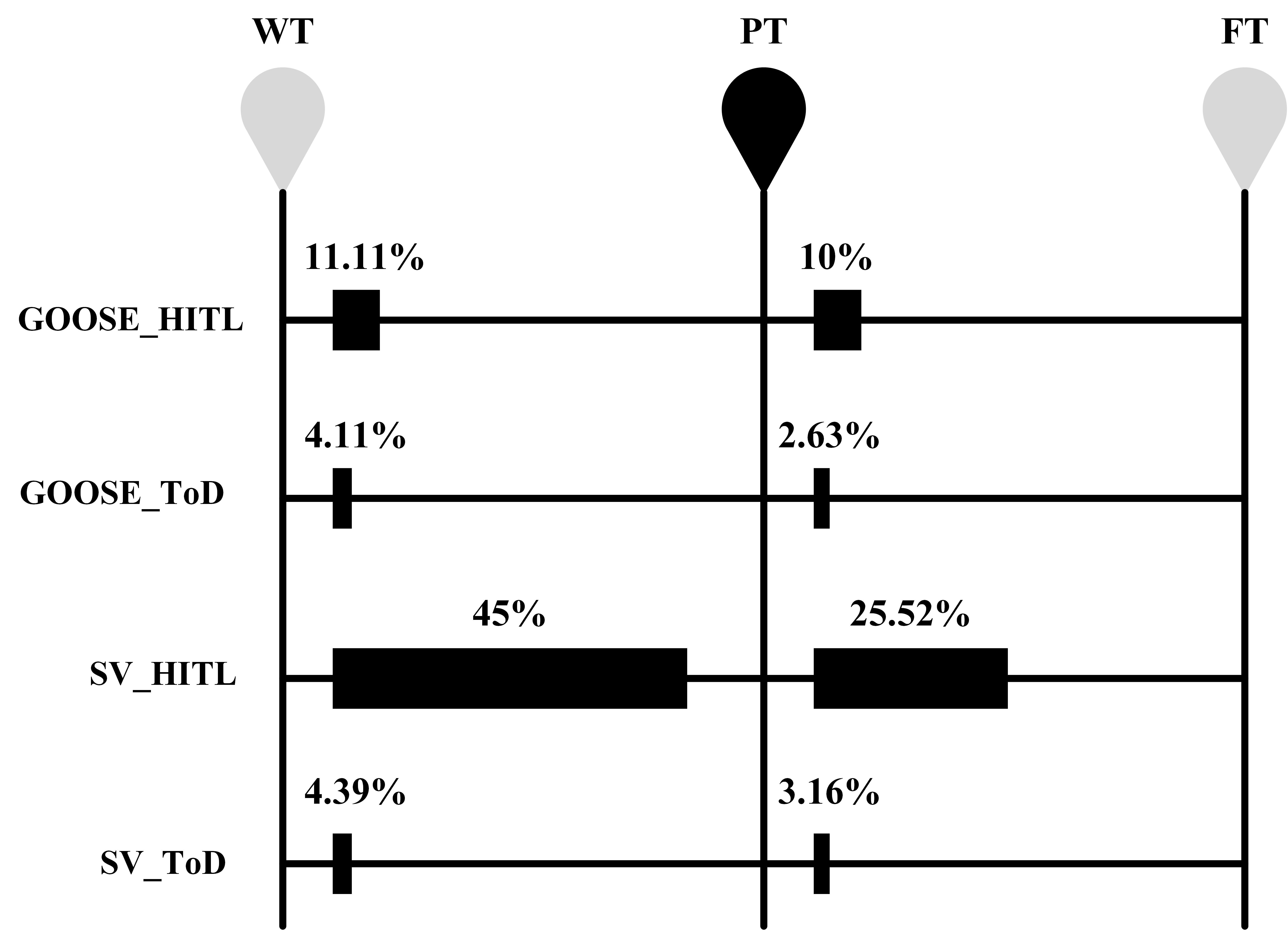}}
\caption{Accuracy metrics for various models based on training levels.}
\label{accuracydifference}
\end{figure}

A comparative analysis is carried out to evaluate the performance of the proposed framework against the HITL process. This evaluation aims to illuminate the potential advantages of the proposed approach, offering insights into its scalability and adaptability in complex scenarios. 
%The subsequent sections present the results of the proposed AD, employing advanced metrics for comparison with the HITL process.
This study utilizes advanced evaluation metrics including informedness, markedness, Matthews correlation coefficient (MCC), and geometric mean (GM), each ranging from $-1$ to $1$, except for the GM, which is between $0$ and $1$, to assess the consistency, decision-making, quality, and balance between normal and abnormal datasets, respectively. In the context of GOOSE/SV datasets, these metrics serve distinct purposes: Informedness measures the model's ability to detect anomaly-indicative patterns, Markedness assesses its proficiency in reducing false positives (FPs) and negatives (FNs), MCC provides a balanced performance measure in scenarios with rare anomalies, and the GM evaluates the model's precision in AD while maintaining low FP rates~\cite{de2022general}. In this application, a value of $1$ for informedness, markedness, MCC, or GM indicates optimal performance, reflecting perfect anomaly detector, crucial for maintaining the accuracy and reliability of digital substation communications. This section endeavors to elucidate the results obtained through the application of advanced metrics across diverse methodologies, classified by their respective training levels (i.e., WT, PT, and FT). To simplify this process and minimize human effort, an image-based framework based on rules was conceptualized and implemented. 
%The choice of the primary analytical tool was driven by the observed limitations in other platforms; specifically, some alternatives demonstrated persistent issues with detailed image analysis, while others, being inherently text-focused, proved ineffective for handling image analysis tasks, making them unsuitable for this purpose. 
The empirical evidence presented in Table~\ref{results} demonstrates the superiority of the proposed framework across all training levels. 
%%%%%%%%%%%%%%%%%%%%%%%%%%%%%%%%%%%%%%%%%%%%%%
\begin{table}[!h]
    \centering
    \fontsize{7pt}{9pt}\selectfont
    \caption{A comparative analysis of AD regarding WT, PT, FT levels.}
    \label{results}
    \begin{tabular}{|c|c|c|c|c|c|c|}
        \hline
        \multicolumn{7}{|c|}{\textbf{GOOSE}}
        \\  \hline
        \multicolumn{1}{|c|}{\textbf{Method}} & \multicolumn{3}{c|}{\textbf{HITL}} & \multicolumn{3}{c|}{\textbf{ToD}} \\ \hline
        \textbf{Metrics} & WT & PT & FT & WT & PT & FT \\ \hline
        \textit{Informedness} & 0.22 & 0.3964 & 0.5709 & 0.825 & 0.8998 & 0.9492 \\ \hline
        \textit{Markedness} & 0.233 & 0.416 & 0.599 & 0.8296 & 0.8998 & 0.9492 \\ \hline
        \textit{MCC} & 0.0247 & 0.2054 & 0.4142 & 0.822 & 0.8997 & 0.9491 \\ \hline
        \textit{Geometric Mean} & 0.5865 & 0.6844 & 0.7784 & 0.9105 & 0.9512 & 0.9746 \\ \hline
        \multicolumn{7}{|c|}{\textbf{SV}}
        \\  \hline
        \multicolumn{1}{|c|}{\textbf{Method}} & \multicolumn{3}{c|}{\textbf{HITL}} & \multicolumn{3}{c|}{\textbf{ToD}} \\ \hline
        \textbf{Metrics} & WT & PT & FT & WT & PT & FT \\ \hline
        \textit{Informedness} & \textbf{0} & 0.5 & 0.8833 & 0.8296 & 0.8968 & 0.9467 \\ \hline
        \textit{Markedness} & \textbf{0} & 0.3836 & 0.7407 & 0.825 & 0.8968 & 0.9467 \\ \hline
        \textit{MCC} & \textbf{0} & 0.3713 & 0.8432 & 0.823 & 0.8966 & 0.9466 \\ \hline
        \textit{Geometric Mean} & 0.5 & 0.7483 & 0.9397 & 0.9143 & 0.9484 & 0.9733 \\ \hline
    \end{tabular}
\end{table}
%%%%%%%%%%%%%%%%%%%%%%%%%%%%%%%%%%%%%%%%%%%%%
By comparison, the HITL process exhibits suboptimal performance, with the first three metrics yielding values of $0$, indicating that its predictive capabilities for positive and negative data rarely surpass random chance. This deficiency manifests in elevated rates of FPs and FNs, substantially compromising the prediction accuracy. An in-depth analysis of Fig.~\ref{MetricAll} reveals the proposed method's unquestionable superiority across all metrics, thereby affirming its reliability in the context of AD for GOOSE/SV datasets.
Moreover, the model's decision-making efficacy when applied to SV dataset is comparable to that of a coin toss, rendering these models unreliable due to the absence of meaningful correlations and the inability to consistently identify correct predictions. The advanced metrics visualized in Fig.~\ref{MetricAll} further confirm the exceptional performance of the GenAI-based framework, particularly at the FT level, where values approaching $1$ underscore its good efficiency, scalability, adaptability, and reliability.
Additionally, Fig.~\ref{accuracydifference} presents a novel comparative framework, investigating various models across different training levels based on accuracy and improvement differentials.
This visualization serves to articulate the incremental percentage gains at each level. Notably, this model exhibits minimal increases when implemented within the proposed framework, suggesting a robust correlation across training levels and demonstrating exceptional AD capabilities even in the absence of training, as evidenced by the negligible performance disparities across levels. Conversely, the HITL process displays the highest incremental percentage, indicating enhanced adaptability in synthesizing new rules. Nevertheless, the integration of HITL methodology further enhances the ToD framework by incorporating human expertise. This combination allows for a refinement of the model's accuracy and responsiveness through continuous feedback mechanisms. As a result, the AD process becomes more precise and dependable, effectively leveraging both automated systems and human insight to identify and respond to anomalies more efficiently.
%%%%%%%%%%%%%%%%%%%%%%%%%%%%%%%%%%%%%%%%
\section{Conclusion and Future Directions} \label{conclusion}
This study introduces a GenAI-based ToD framework for an efficient and reliable AD in IEC61850-based multicast messages. The framework's scalability and adaptability are validated through comparative analysis with HITL process, employing advanced metrics to assess reliability and correlation capabilities—aspects previously overlooked in smart grids. Also, this GenAI-based methodology demonstrates acceptable performance across all metrics. Future research directions will include integrating a self-learning component to expand the applicability to other messages, such as MMS, and incorporating natural language processing (NLP) metrics to enhance the quality of GenAI-generated outputs.
%%%%%%%%%%%%%%%%%%%%%%%%%%%%%%%%%%%
\bibliographystyle{IEEEtran}
\bibliography{IEEEabrv,ref}

\end{document}